# Enhancing Building Energy Efficiency through Advanced Sizing and Dispatch Methods for Energy Storage

**Min Gyung Yu, PhD**

**Xu Ma, PhD**

**Bowen Huang, PhD**

**Karthik Devaprasad**
*Associate member*

**Fredericka Brown, PhD**

**Di Wu, PhD**

**ABSTRACT HEADING**

*Energy storage and electrification of buildings hold great potential for future decarbonized energy systems. However, there are several technical and economic barriers that prevent large-scale adoption and integration of energy storage in buildings. These barriers include integration with building control systems, high capital costs, and the necessity to identify and quantify value streams for different stakeholders. To overcome these obstacles, it is crucial to develop advanced sizing and dispatch methods to assist planning and operational decision-making for integrating energy storage in buildings. This work develops a simple and flexible optimal sizing and dispatch framework for thermal energy storage (TES) and battery energy storage (BES) systems in large-scale office buildings. The optimal sizes of TES, BES, as well as other building assets are determined in a joint manner instead of sequentially to avoid sub-optimal solutions. The solution is determined considering both capital costs in optimal sizing and operational benefits in optimal dispatch. With the optimally sized systems, we implemented real-time operation using the model-based control (MPC), facilitating the effective and efficient management of energy resources. Comprehensive assessments are performed using simulation studies to quantify potential energy and economic benefits by different utility tariffs and climate locations, to improve our understanding of the techno-economic performance of different TES and BES systems, and to identify barriers to adopting energy storage for buildings. Finally, the proposed framework will provide guidance to a broad range of stakeholders to properly design energy storage in buildings and maximize potential benefits, thereby advancing affordable building energy storage deployment and helping accelerate the transition towards a cleaner and more equitable energy economy.*

## INTRODUCTION

The increasing impact of climate change requires immediate action to achieve a net-zero greenhouse gas emissions economy. While renewable energy sources, such as solar and wind, are essential to achieving decarbonization goals, their intermittent characteristics bring challenges to utilizing clean energy efficiently and balancing supply and demand in grid operation. Energy storage systems, such as battery energy storage (BES) and thermal energy storage (TES) systems, provide promising solutions to facilitate renewable energy sources and enhance energy efficiency. Accordingly, the global energy storage market is estimated to grow more over the forecast period (James, et al. 2021) (Ralon, et al. 2017).

**Yu et al.** are research engineers at Pacific Northwest National Laboratory, Richland, WA.

With the growing adoption of building energy storage, there is a pressing need for technical support to utilize these energy storage systems effectively and efficiently. Building energy storage has been extensively studied to improve overall system efficiency and lower operation costs as reviewed in (Heier, Bales and Martin 2015). On top of that, the optimal sizing problem of the energy storage systems is crucial due to large capital costs and space limits. However, there are a few limitations in previous research for advancement in methodology and analysis. First, the development of a scalable and flexible approach is needed. Many previous studies have utilized metaheuristic algorithms, such as the genetic algorithm and particle swarm optimization (PSO), for the energy system optimization problem with TES and BES (Ikeda and Ooka 2015). One previous study solved the complicated optimal problem of the combined usage of TES and BES for a residential building with photovoltaic (PV) panels using the PSO algorithm to minimize the life cycle costs (Baniasadi, et al. 2020). Although this research validated the effectiveness of the method to support valuable design choices, the metaheuristic algorithm used in previous research has limitations in finding the global optimum solution for complex and nonlinear problems. As the problem becomes more complicated, mathematical optimization techniques are advantageous over metaheuristics, as they can handle large and complex systems efficiently and can provide guaranteed optimal solutions. In this context, a prior study formulated a mixed-integer linear programming model to address the optimal dispatch problem of building BES and TES (Niu, et al. 2019). The research explored the flexibility potential; however, optimal sizing was not investigated in this work. Secondly, there is a growing need to quantitatively assess potential benefits at the regional level under various scenarios. This comprehensive evaluation should consider various factors, including building types, climate regions, and utility rate tariffs. Addressing these challenges is crucial for maximizing the benefits of energy storage technologies and promoting their broader integration into sustainable energy systems.

In this research, we propose a flexible framework that considers both BES and TES systems for a building, using mixed-integer optimization programming to efficiently utilize energy resources while meeting the building's energy demand. Our approach also incorporates the impact of the power market which has different electricity pricing and demand charge pricing in different regions, to provide a more realistic and practical solution. This research paper aims to highlight the significance of optimal sizing and dispatch strategies in accelerating decarbonization efforts and mitigating the impacts of climate change.

**BUILDING ENERGY MODEL**

In this section, the building energy system models required for the optimal sizing and dispatch of BES and TES within building are described. TES is modeled with thermal energy systems (e.g., chiller, heat pump, boiler) to meet building thermal demands. The thermal systems are designed not only to provide thermal energy to rooms but also to store excess thermal energy within TES. The energy consumptions of the thermal systems are accounted for in the overall building electric demand, which we will describe with the BES model. Note that the proposed framework can be applied to both heating and cooling operation, but this paper specifically focuses on cooling mode. Figure 1 provides an overview of building energy demands and energy systems considered in this paper. An ice-storage is used as TES, and chillers are modeled as thermal systems for TES operation and meeting the cooling demand. The chiller operation affects the overall building electricity consumption, and it can be provided directly from the grid or BES operation.

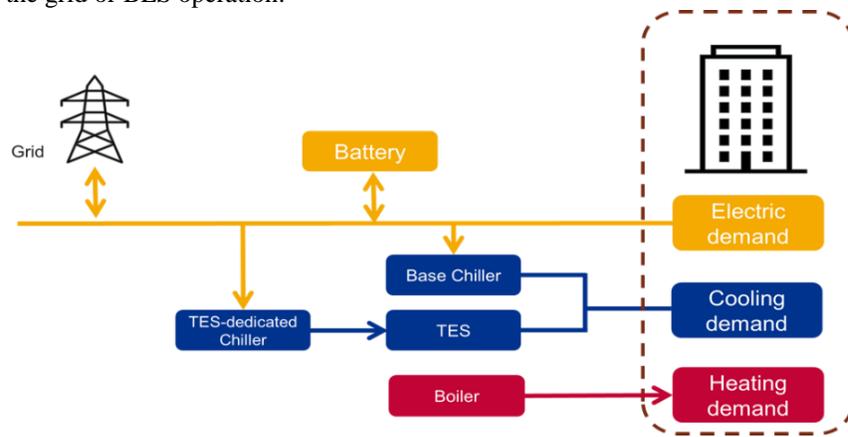

**Figure 1** Overview of building energy demands and energy systems.

## Building Energy System

The primary cooling energy system in this building model includes a base chiller, a TES-dedicated chiller, and a TES with two chilled water loops (See Figure 1). One chilled water loop is designed for TES charging with a dedicated chiller, while the other chilled water loop includes both the base chiller and TES to meet cooling requirements. During the TES charging, the TES-dedicated chiller functions to store the thermal energy in TES. When TES is discharging, the cooling demand is met by thermal energy supplied through both TES discharging and base chiller operation.

**Chiller model.** Our chiller model is developed based on the electric chiller model from EnergyPlus, which encompasses all the features of the DOE-2.1 chiller model (U.S. Department of Energy n.d.). The chiller model leverages three performance curves, which are: i) Cooling Capacity Function of Temperature Curve ($\Psi_1$), ii) Energy Input to Cooling Output Ratio Function of Temperature Curve ($\Psi_2$), and iii) Energy Input to Cooling Output Ratio Function of Part Load Ratio Curve ($\Psi_3$). The chiller's part-load ratio (PLR) is defined as the ratio of the cooling load ($Q_{load}$) to the available cooling capacity ($Q_{avail}$), where $Q_{avail}$ will be adjusted with the performance by chilled water supply setpoint temperature ($T_{cw,ls}$) and condenser leaving temperature ($T_{cond,l}$). The electrical power consumption for the chiller compressor is then calculated by multiplying the reciprocal of the reference coefficient of performance ($COP_{ref}$) and chiller capacity ($C_{chiller}$) at a given operating condition with the performance curves. This model simulates the thermal performance of the chiller and the power consumption of the compressor. Mathematically, the chiller model can be written as:

$$\Psi_1 = a_0 + a_1 T_{cw,ls} + a_2 T_{cw,ls}^2 + a_3 T_{cond,l} + a_4 T_{cond,l}^2 + a_5 T_{cw,ls} T_{cond,l} \tag{1a}$$

$$\Psi_2 = b_0 + b_1 T_{cw,ls} + b_2 T_{cw,ls}^2 + b_3 T_{cond,l} + b_4 T_{cond,l}^2 + b_5 T_{cw,ls} T_{cond,l} \tag{1b}$$

$$\Psi_3 = c_0 + c_1 PLR + c_2 PLR^2 \tag{1c}$$

$$Q_{avail} = C_{chiller} \Psi_1 \tag{1d}$$

$$PLR = Q_{load}/Q_{avail} \tag{1e}$$

$$P_{chiller} = Q_{avail} \frac{1}{COP_{ref}} (\Psi_2)(\Psi_3) \tag{1f}$$

**TES model.** This work considers an ice-storage as TES. The TES dynamics can be expressed as a function of the charging and discharging rates. The state-of-charge (SOC) level of TES at time $t$ ($l_{tes}^t$) is used to denote the stored energy as a percentage of the TES capacity ($Q_{tes}$). The charging rate ($q_{ch}^t$) should be lower than the TES-dedicated chiller capacity ($C_{TES-chiller}$) and maximum charging rate ($\overline{q_{ch}}(l)$); and the discharging rate ($q_{dis}^t$) should be lower than the cooling demand ($Q_{cool}$), maximum discharging rate ($\overline{q_{dis}}(l)$). Both charging and discharging rates should also be limited by the SOC level constraints ($\underline{L_{tes}}, \overline{L_{tes}}$). Mathematically, the TES dynamics can be expressed as follows:

$$l_{tes}^t = l_{tes}^{t-1} + \frac{q_{ch}^t - q_{dis}^t}{Q_{tes}} \tag{2a}$$

$$\underline{L_{tes}} \leq l_{tes}^t \leq \overline{L_{tes}} \tag{2b}$$

$$0 \leq q_{ch}^t \leq \min\left(C_{TES-chiller}, (\overline{L_{tes}} - l^{t-1})Q_{tes}, \overline{q_{ch}}(l)\right) \tag{2c}$$

$$0 \leq q_{dis}^t \leq \min\left(Q_{cool}, (l^{t-1} - \underline{L_{tes}})Q_{tes}, \overline{q_{dis}}(l)\right) \tag{2d}$$

**BES model.** The BES dynamics can also be expressed as a function of the charging and discharging rates. The SOC level of BES at time $t$ ($k_{bes}^t$) is used to denote the stored energy as a percentage of the BES capacity ($C_{bes}$). The charging and discharging rates ($p_{bes,ch}^t, p_{bes,dis}^t$) are adjusted by efficiency ($\eta_{bes,ch}, \eta_{bes,dis}$) and should be lower than maximum rates ($\overline{P_{bes}}$). Additionally, both charging and discharging rates should also be limited by the SOC level constraints ($\underline{K_{bes}}, \overline{K_{bes}}$). The BES dynamics can be expressed as:

$$k_{bes}^t = k_{bes}^{t-1} + \left(p_{bes,ch}^t \cdot \eta_{bes,ch} - p_{bes,dis}^t/\eta_{bes,dis}\right)/C_{bes} \tag{3a}$$

$$\underline{K_{bes}} \leq k_{bes}^t \leq \overline{K_{bes}} \tag{3b}$$

$$0 \leq p_{bes,ch}^t \leq \overline{P_{bes}} \text{ and } 0 \leq p_{bes,dis}^t \leq \overline{P_{bes}} \tag{3c}$$

$$P_{bes}^t = p_{bes,ch}^t - p_{bes,dis}^t \tag{3d}$$

**Building Energy Demand**

Building cooling demand is to maintain the room temperature at its specific setpoint temperature considering heat gain/loss, and the cooling system operates to meet it. To guarantee indoor thermal comfort, the base chiller ($Q_{avail,base}^t$) and TES discharging must collectively satisfy the cooling demand:

$$Q_{avail,base}^t + q_{dis}^t \geq Q_{load}^t \tag{4}$$

Building electric demand is the sum of the total HVAC electric load and other lighting and plug loads. It can be further categorized into flexible and non-flexible components, and the flexible load is related to the BES and TES operation. In this work, flexible loads are charging and discharging power of the battery ($P_{bes}$) and power consumption of the TES-dedicated chiller and base chiller ($P_{chiller}$) integrated with TES operation. There are other HVAC power consumptions from cooling tower and water pump; however, these are not involved in this work due to its small fraction of the total (Drees and Braun 1996). The lighting, plug electric loads, and fan power of the distribution side are categorized as non-flexible loads ($P_{non}$). Overall, the actual total building electric ($P_{total}$) is estimated as follows:

$$P_{total} = P_{non} + P_{chiller} + P_{bes} \tag{5}$$

**OPTIMAL SIZING AND DISPATCH FRAMEWORK**

The optimization framework associated with the sizing and dispatch problem is illustrated in Figure 2. To begin with, we consider the optimal dispatch problem as a model predictive control (MPC) problem. We use a model interval denoted by $\Delta t$ and a control interval denoted by $\Delta T = \overline{M}\Delta t$, so that the prediction horizon of the MPC can be expressed as $T = \overline{N}\Delta T = \overline{N}\,\overline{M}\Delta t$, where $\overline{N}$ and $\overline{M}$ are positive integers. In this work, we set $\overline{N} = K$ and $\overline{M} = 1$. A typical MPC problem can be implemented as follows: At time $t = 0$, the MPC computes a sequence of optimal control actions, denoted by $\{u_0, u_{\Delta T}, \cdots, u_{(K-1)\Delta T}\}$, for the period of the first prediction horizon $[0, K\Delta T)$. Only the first control action $u_0$ is applied to the system at $t = 0$. Then, the same control process described above will be repeated for every time instant $t = \Delta T, 2\Delta T, 3\Delta T, \cdots$, until the final control time is reached. A general optimal dispatch problem associated with the MPC can be represented as:

$$\min_{U} J = \sum_{t=1}^{K} \lambda^t P_{total}^t + \nu^m P_{peak}^m \tag{6a}$$

subject to

$$P_{peak}^m \geq P_{total}^t, \; \forall t \in \{t | month(t) = m\} \tag{6b}$$

Chiller model (1), TES model (2), BES model (3), Building Energy Demand (4), (5) (6c)

where $J$ represents the objective function to minimize the total operational costs, including both energy and demand charges, and $P_{peak}^m$ denotes the peak demand within month $m$. In the optimal dispatch problem (6), the decision variables are $U = \{q_{ch}^t, q_{dis}^t, l_{tes}^t, p_{bes,ch}^t, p_{bes,dis}^t, k_{bes}^t, P_{chiller}, P_{bes}^t, P_{total}^t, P_{peak}^m\}$. We also use $\lambda^t$ and $\nu^m$ to denote the energy charge at time $t$ and the demand charge for month $m$, respectively.

In the optimal sizing problem, the decision variables also include the base chiller capacity ($C_{chiller}$) and the TES-dedicated chiller capacity ($C_{TES-chiller}$), the capacity of TES ($Q_{tes}$), the power capacity ($\overline{P_{bes}}$) and energy capacity of BES ($C_{bes}$). Hence, the sizing problem can be solved for the project lifespan by minimizing the present value costs of the whole project:

$$\min_U J = \sum_{t=1}^K \lambda^t P_{total}^t + \nu^m P_{peak}^m + \alpha_1 C_{chiller} + \alpha_2 C_{TES-chiller} + \alpha_3 Q_{tes} + \alpha_4 \overline{P_{bes}} + \alpha_5 C_{bes} \quad (7)$$

where coefficients $\alpha_1, \cdots, \alpha_5$ represent the installed cost prices for TES and BES. The present value costs expressed in (7) include the system's capital costs, as well as the present value operational costs.

In real practice, we first solve the optimal sizing problem off-line, and then pass the optimal sizing results as inputs into the MPC (optimal dispatch) solver. The MPC will be solved repeatedly at the optimization stage, and the MPC solver will also keep communicating with the EnergyPlus co-simulation module so that the solution for the next time interval will be used as TES/BES control signals in the EnergyPlus model.

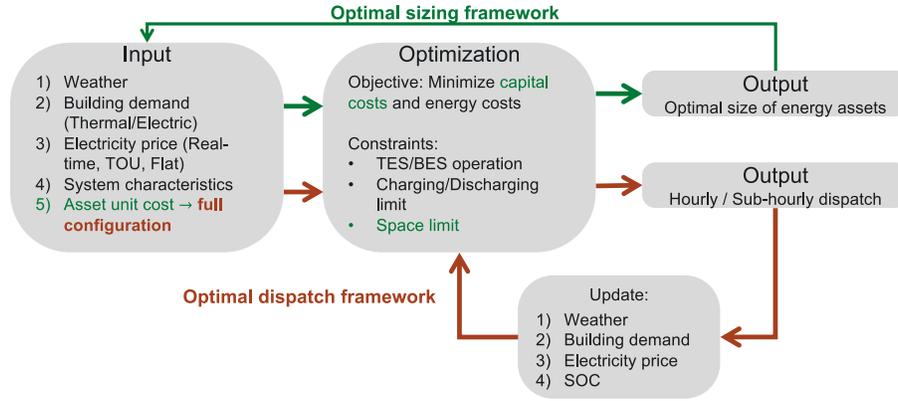

**Figure 2** Optimal sizing and dispatch framework for TES and BES.

## CASE STUDY

The proposed framework can be applied to any building type for optimal sizing and economic analysis. This study selected DOE prototype large office buildings (U.S. Department of Energy 2020) to showcase best practices in energy management and efficiency, given their status as significant energy consumers. The building model has a data center in the basement, where cooling is provided by a water-source DX cooling coil, while the rest of the building is cooled using two chillers. To assess the potential benefits of building energy storage depending on climate location and utility tariffs, we chose three locations in different climate zones: California, Boston, and Texas. These locations were selected based on their distinct geographic, climatic, and economic characteristics. This work leveraged EnergyPlus, a detailed building energy simulation tool, to estimate the thermal and electric demand of a building and to implement MPC in a co-simulation setup.

Figure 3 shows the one-year weather profiles and the corresponding electric and cooling demand of three different locations, excluding electric demand from the data center, as simulated using EnergyPlus. California generally has mild temperatures throughout the year. Boston's weather is characterized by cold winters and humid summers, whereas Texas tends to have hot summers with milder and moderate winters. Accordingly, the peak cooling demand was approximately 1200 kW in California, while it was around 2780 kW and 3100 kW in Boston and Texas, respectively. Peak electric demand was measured at around 1200 kW, 1400 kW and 1530 kW in California, Boston, Texas, respectively.

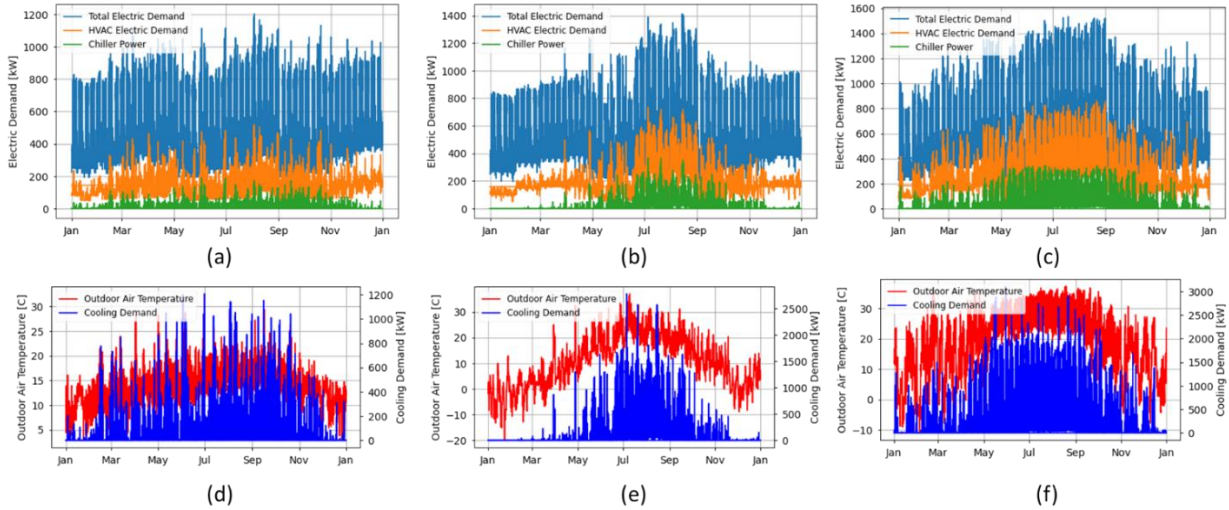

**Figure 3** (a) Electric demand and (d) outdoor air temperature and cooling demand of California, (b) Electric demand and (e) outdoor air temperature and cooling demand of Boston, (c) Electric demand and (f) outdoor air temperature and cooling demand of Texas for over a one-year period.

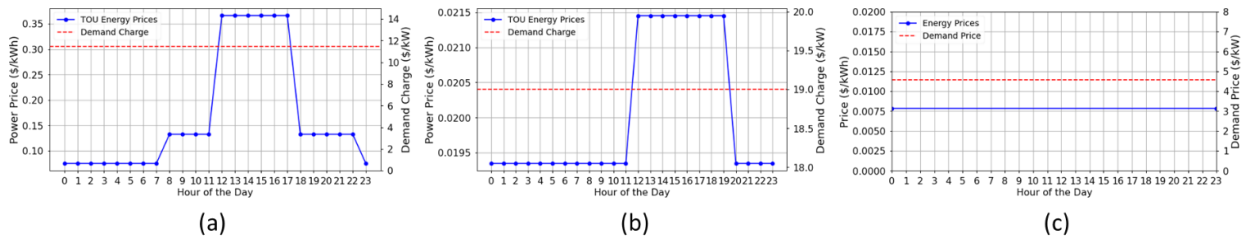

**Figure 4** Electric utility rate tariffs in (a) California, (b) Boston, and (c) Texas for summer season weekday schedule.

**Table 1** Sizing parameters for unit prices, capacity and maximum space limits of the energy assets.

|  | Unit Price | Max. Capacity | Max. Space |
|---|---|---|---|
| Chillers | $120/kW | 10000 kW |  |
| TES | $40/kWh | 10000 kWh | 500 m$^2$ |
| BES | $355/kWh | 10000 kWh | 500 m$^2$ |
| BES power | $153/kW | 10000 kW |  |

In Table 2, the optimal sizes of energy assets are determined for three different locations using the electric utility rate tariffs (See Figure 4) taken from OpenEI's Utility Rate Database (OpenEI n.d.) to characterize how the economic benefits of building energy storages (TES and BES) vary. Note that we selected time-of-use (TOU) rate tariffs for both California and Boston, because TOU rates encourage customers to reduce their energy consumption during peak hours when electricity costs are higher and increase consumption during off-peak hours when rates are lower. As for Texas case, we did not find any TOU rate tariff with the selected zip codes that we used in Database and thus we considered a flat rate tariff instead. The time frame to be studied is 20 years and the discount rate is assumed to be 5%. The unit price taken from Energy Storage Cost and Performance Database (Pacific Northwest National Laboratory n.d.) and physical constraints (e.g., capacity, space) are described in Table 1. The assumed charging and discharging efficiencies (including both battery and inverter) are 0.93.

**Table 2 Optimal size of energy assets for large office building in different locations**

|  | California | Boston | Texas |
|---|---|---|---|
| Base chiller (kW) | 900 | 2300 | 2700 |
| TES dedicated chiller (kW) | 200 | 300 | 100 |
| TES (kWh) | 3300 | 4400 | 1200 |
| BES (kWh) | 8000 | 600 | 0 |
| BES power (kW) | 1000 | 500 | 0 |

In California, it proves highly cost-effective to invest in a large TES and BES. The base chiller's capacity is reduced by more than 25% of the peak cooling demand. The optimal sizing of BES is more than 6 times the peak electric demand, and the optimal discharging duration of BES is about 8 hours. In contrast, the optimal sizing solution in Boston involves a larger TES with a smaller BES. Using the TES, the base chiller's capacity is reduced by around 17% compared to the peak cooling demand. In Texas, limited cost-effectiveness of BES makes a small TES a more economically viable option, resulting in a 13% reduction in the base chiller size compared to the peak cooling demand. It's important to note that the optimal sizing solutions are primarily determined by the utility tariffs in each location. California has higher electricity rates, particularly with its dynamic TOU power pricing structure. It has significant fluctuations in power prices throughout the day, with prices during 12 PM to 6 PM roughly 3 times higher than off-peak hours. In Boston, while electricity prices are relatively lower, the demand charge price is around $19/kW, which is the highest among the three locations. In contrast, Texas maintains a flat power price with the lowest demand charge price among the three locations. As a result, the sizing calculations propose limited investment in building energy storage systems. Please note that these obtained optimal sizes should only be used as guidance. In practice, due to the limited available models from manufacturers, the exact optimal sizes may not be an option for deployment and commercial sizes around the optimum need to be explored and selected.

To assess the benefits of building energy storage assets with the proposed optimal sizing and dispatch, we compare the performance with the baseline which is the original prototype building model without TES and BES. The equipment in the base case is sized using EnergyPlus, taking into account the building configuration and external design conditions, while the energy assets in the optimal case is designed leveraging the proposed optimal sizing framework (See Table 2). Table 3 provides a summary of the energy and cost benefits of the optimal building energy storage over one-week operation. By utilizing TES and BES, we observe more energy consumption compared to the baseline, ranging from 1.41% to 4.34%. As shown in Figure 5, the SOC levels of TES and BES increase as they store excess energy when the energy prices and demands are low, and they decrease as they use the stored energy when the energy prices and demands are high. However, the peak demand is reduced by 2.15 to 9.2%. The savings in the electricity bill vary significantly, and this is affected by the utility tariff structure and the efficiencies of TES and BES. The optimization engine captures the trade-offs between savings in energy and demand charges, and eventually maximizes the total savings in the electricity bill. Under circumstances where the energy loss from energy shifting can not be fully compensated by the different hourly energy rates (e.g., Boston and Texas), the energy charges with TES and BES may be even higher than the base case without TES and BES. In this case, the optimization engines capture the savings in demand charges instead. The demand charge savings are 9.21% in California, 3.8% in Boston, and 2.15% in Texas. Eventually, the total savings in electricity bill are 5.9% in California, 2.9% in Boston, and 1.5% in Texas. Overall, the capability of lowering peak demand from TES and BES provides a great opportunity to cut the electricity bill.

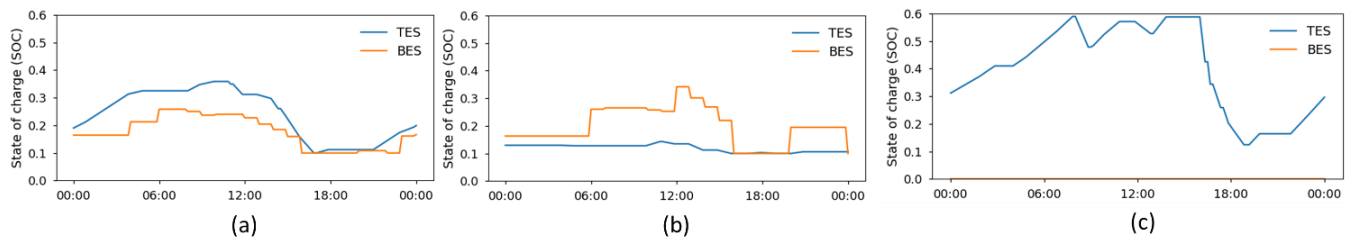

**Figure 5** State of charge (SOC) trends of TES and BES in (a) California, (b) Boston, and (c) Texas.

**Table 3 Energy and cost benefits with energy assets specified in Table 2 over one-week operation.**

|  |  | California | Boston | Texas |
|---|---|---|---|---|
| Energy usage (total, peak) | w/o TES,BES | 195MWh, 1570kW | 220MWh, 1815kW | 244MWh, 1890kW |
|  | w/ TES,BES | 203MWh, 1425kW | 230MWh, 1746kW | 247MWh, 1850kW |
|  | saving | -8MWh, 145kW | -10MWh, 69kW | -3MWh, 40kW |
| Energy charge | w/o TES,BES | $36786 | $4439 | $1915 |
|  | w/ TES,BES | $35200 | $4628 | $1942 |
|  | saving | $1586 (4.31%) | $ (-189) (-4.25%) | $(-27) (-1.41%) |
| Demand charge | w/o TES,BES | $17975 | $34487 | $8623 |
|  | w/ TES,BES | $16319 | $33171 | $8438 |
|  | saving | $1656 (9.21%) | $1316 (3.8%) | $185 (2.15%) |
| Total operation cost saving |  | $3242(5.9%) | $1127(2.9%) | $158 (1.5%) |

## CONCLUSION

This paper presented an optimal sizing and dispatch framework for building energy storage systems and assessed the economic benefits for three climate locations using different utility tariffs. Comprehensive analyses were carried out on the DOE prototype large office buildings using the proposed methods. As expected, the optimal building energy storage size varies with climate locations and electric utility rate tariff, and capital costs. It was found that the utility tariff with higher energy price and demand charge significantly impacted the sizing of building energy storage systems, particularly BES, due to its higher capital costs. One interesting future work is to evaluate the environmental benefits of the proposed framework and demonstrate the effectiveness with various use case scenarios (e.g., building types with diverse load profiles). We also plan to develop a design guideline to determine the optimal sizes of BES and TES for large-scale adoption.

## ACKNOWLEDGMENTS

This research was supported by the Energy Mission Seed Investment, under the Laboratory Directed Research and Development Program at Pacific Northwest National Laboratory (PNNL). PNNL is a multi-program national laboratory operated for the U.S. Department of Energy by Battelle Memorial Institute under Contract No. DE-AC05-76RL01830